\documentclass[aps,prc,twocolumn,fleqn,floatfix,twoside]{revtex4}

\usepackage[dvipdfm]{graphicx}% Include figure files
\usepackage{amsmath,amssymb,amsfonts}

\usepackage{color}
\definecolor{joerg}{rgb}{1.0,0.0,0.0}
\newcommand{\jr}[1]{#1}

\begin{document}

\title{Jet-quenching in a 3D hydrodynamic medium}

\author{Thorsten Renk}
\affiliation{Department of Physics, PO Box 35 FIN-40014 University of Jyv\"{a}skyl\"{a}, Finland}
\affiliation{Helsinki Institute of Physics, PO Box 64 FIN-00014 University of Helsinki, Finland}

\author{J\"org Ruppert}
\affiliation{Department of Physics, McGill University, H3A 2T8, Montreal, Quebec, Canada}

\author{Chiho Nonaka}
\affiliation{Department of Physics, Nagoya University, Nagoya 464-8602, Japan}

\author{Steffen A. Bass}
\affiliation{Department of Physics, Duke University, Durham, NC 27708, USA}

\date{\today}

\begin{abstract}
We study the radiative energy loss of hard partons in a soft medium in the multiple soft scattering approximation. The soft medium is described by a 3D hydrodynamical model and we treat the averaging over all possible parton paths through the medium without approximation. While the nuclear suppression factor $R_{AA}$ does not reflect the high quality of the medium description (except in a reduced systematic uncertainty in extracting the quenching power of the medium), the hydrodynamical model also allows to study different centralities and in particular the angular variation of $R_{AA}$ with respect to the reaction plane, allowing for a controlled variation of the in-medium path-length. We study the angular dependence of $R_{AA}$ for different centralities, discuss the influence of hydrodynamical expansion and flow and comment on the comparison with preliminary data.
\end{abstract}
\maketitle

Experiments at the 
Relativistic Heavy Ion Collider (RHIC) 
 have established a significant suppression
of high-$p_T$ hadrons produced in central A+A collisions compared to those
produced in peripheral A+A or binary scaled p+p reactions, indicating
a strong nuclear medium effect \cite{Adcox:2001jp,Adler:2002xw}. 
The origin of this phenomenon, commonly referred to as {\em jet-quenching}, can
be understood in the following way:
during the early pre-equilibrium stage of the relativistic heavy-ion
collision, scattering of partons which leads to the formation
of deconfined quark-gluon matter often occurs with large momentum transfers 
which leads to the formation of two back-to-back hard partons.  These
traverse the dense medium, losing energy and finally
fragment into hadrons which are observed by the experiments.
Within the framework of perturbative QCD, the leading process of energy 
loss of a fast parton is gluon radiation induced by elastic collisions 
of the leading parton or the radiated gluon with color charges in the
quasi-thermal medium \cite{Gyulassy:1993hr,Baier:1996kr,Zakharov:1997uu}.

Over the past two years, a large amount of 
jet-quenching related experimental data has become available 
including but not limited to
the nuclear modification factor $R_{AA}$, the elliptic
flow $v_2$ at high $p_T$ (as a measure of the
azimuthal anisotropy of the jet cross section) and
a whole array of high $p_T$ hadron-hadron correlations
Computations of such jet modifications have acquired a
certain level sophistication regarding the
incorporation of the partonic processes involved.
However, most of these calculations have been utilizing
over-simplified models for the underlying soft medium,
e.g. assuming a simple density distribution and its
variation with time. Even in more elaborate setups, most jet
quenching calculations assume merely a one-dimensional Bjorken
expansion.

The availability of a three-dimensional hydrodynamic evolution
code \cite{Nonaka:2006yn} allows
for a much more detailed study of jet interactions in a
longitudinally and transversely expanding medium.
The variation of the gluon density in such a medium
is very different from  that in a simple Bjorken expansion.
A previous calculation in this direction \cite{Hirano:2002sc,Hirano:2003pw}
estimated the effects of 3-D expansion on the $R_{AA}$.
However, this approach treated the
energy loss of jets in a rather simplified and heuristic manner.
Here, we shall perform a detailed investigation of the
modification  of jets in a  three dimensionally
expanding medium within a formalism of \cite{Salgado:2003gb}.

Relativistic Fluid Dynamics (RFD, see e.g.
\cite{Bjorken:1982qr,Clare:1986qj,Dumitru:1998es})
is ideally suited for the high-density phase of heavy-ion reactions
at RHIC,
but breaks down in the later, dilute, stages of the
reaction when the mean free paths of the hadrons become
large and flavor degrees of freedom
are important. The biggest advantage  of  RFD
is that it directly incorporates an equation of state as input
and thus is so far the only dynamical model in which a phase
transition can
explicitly be incorporated. 
Starting point for a RFD calculation is the
relativistic hydrodynamic equation
\begin{equation}
\partial_\mu T^{\mu \nu} = 0,
\label{Eq-rhydro}
\end{equation}
where $T^{\mu \nu}$ is the energy momentum tensor which is given by
\begin{equation}
T^{\mu \nu}=(\epsilon + p) U^{\mu} U^{\nu} - p g^{\mu \nu}.
\end{equation}
Here $\epsilon$, $p$, $U$ and $g^{\mu \nu}$ are energy density,
pressure, four velocity and metric tensor, respectively.
The relativistic hydrodynamic equation Eq.\ (\ref{Eq-rhydro})
is solved numerically using baryon number $n_B$ conservation
\begin{equation}
\partial_\mu (n_B (T,\mu) U^\mu)=0.
\end{equation}
as a constraint and closing the resulting set of partial
differential equations by specifying an equation of state (EoS):
$\epsilon = \epsilon(p)$.
In the ideal fluid approximation
(i.e. neglecting off-equilibrium effects) and once the initial conditions
for the calculation have been fixed,
the EoS is the {\em only}
input to the equations of motion and relates directly to
properties
of the matter under consideration. Ideally, either the initial conditions or the
EoS should be determined beforehand by an ab-initio calculation (e.g. for the EoS via
a lattice-gauge calculation), in which case a fit to the data would allow
for the determination of the remaining quantity.
Our particular RFD implementation
utilizes a Lagrangian mesh and 
light-cone coordinates $(\tau,x,y,\eta)$ ($\tau=\sqrt{t^2-z^2}$),
in order to optimize the model for ultra-relativistic regime
of heavy collisions at RHIC.

We assume that hydrodynamic expansion starts at
$\tau_0=0.6$ fm. Initial energy density and
baryon number density are parameterized by
\begin{eqnarray}
\epsilon(x,y,\eta)& =& \epsilon_{\rm max}W(x,y;b)H(\eta),
\nonumber \\
n_B(x,y,\eta)& = & n_{B{\rm max}}W(x,y;b)H(\eta),
\end{eqnarray}
where $b$ and  $\epsilon_{\rm max}$ ($n_{B{\rm max}}$) are
the impact parameter and the maximum value of energy density
(baryon number density), respectively.
$W(x,y;b)$ is given by a combination of wounded nuclear model and
binary collision model \cite{Kolb:2001qz} and  $H(\eta)$ is given
by $\displaystyle
H(\eta)=\exp \left [ - (|\eta|-\eta_0)^2/2 \sigma_\eta^2 \cdot
\theta ( |\eta| - \eta_0 ) \right ]$.
RFD has been very successful in describing single soft matter
properties at RHIC, especially
collective flow effects and particle spectra 
\cite{Kolb:2003dz,Huovinen:2003fa,Hirano:2002hv,Nonaka:2006yn}.
All parameters of our hydrodynamic evolution \cite{Nonaka:2006yn}
have been fixed
by a fit to the soft sector (elliptic flow, pseudo-rapidity
distributions and low-$p_T$ single particle spectra), therefore
providing us with a fully determined 
medium evolution for the hard probes to propagate through.

Let us now discuss the treatment of partons propagating through the medium: 
our calculation follows the BDMPS formalism for radiative energy loss 
\cite{Baier:1996sk} using quenching weights as introduced by
Salgado and Wiedemann \cite{Salgado:2002cd,Salgado:2003gb}.
The probability density $P(x_0, y_0)$ for finding a hard vertex at the 
transverse position ${\bf r_0} = (x_0,y_0)$ and impact 
parameter ${\bf b}$ is given by the product of the nuclear profile functions as
\begin{equation}
\label{E-Profile}
P(x_0,y_0) = \frac{T_{A}({\bf r_0 + b/2}) T_A(\bf r_0 - b/2)}{T_{AA}({\bf b})},
\end{equation}
where the thickness function is given in terms of Woods-Saxon the nuclear density
$\rho_{A}({\bf r},z)$ as $T_{A}({\bf r})=\int dz \rho_{(A}({\bf r},z)$.

If we call the angle between outgoing parton and the reaction plane $\phi$, 
the path of a given parton through the medium $\xi(\tau)$ is specified 
by $({\bf r_0}, \phi)$ and we can compute the energy loss 
probability $P(\Delta E)_{path}$ for this path. We do this by 
evaluating the line integrals
\begin{equation}
\label{E-omega}
\omega_c({\bf r_0}, \phi) = \int_0^\infty \negthickspace d \xi \xi \hat{q}(\xi) \quad  \text{and} \quad \langle\hat{q}L\rangle ({\bf r_0}, \phi) = \int_0^\infty \negthickspace d \xi \hat{q}(\xi)
\end{equation}
along the path where we assume the relation
\begin{equation}
\label{E-qhat}
\hat{q}(\xi) = K \cdot 2 \cdot \epsilon^{3/4}(\xi)
\end{equation}
between the local transport coefficient $\hat{q}(\xi)$ (specifying 
the quenching power of the medium) and energy density $\epsilon$.
Here, $\omega_c$ is the characteristic gluon frequency, setting the scale of the energy loss probability distribution and $\langle \hat{q} L\rangle$ is a measure of the path-length, weighted by the local quenching power.
We view 
the parameter $K$ as a tool to account for the uncertainty in the selection of $\alpha_s$ and possible non-perturbative effects increasing the quenching power of the medium (see discussion in \cite{Correlations}).

Using the numerical results of \cite{Salgado:2003gb}, we obtain $P(\Delta E)_{path}$ 
for $\omega_c$ and $R=2\omega_c^2/\langle\hat{q}L\rangle$ as a function of jet production vertex and the angle $\phi$ (here $R$ is a dimensionless quantity needed as input for the energy loss probability distributions as defined in \cite{Salgado:2003gb}).
\jr{The energy loss probability  $P(\Delta E)_{path}$ is derived in the limit of infinite parton energy  \cite{Salgado:2003gb}. In order to account for the finite energy of the partons we truncate $P(\Delta E)$ 
at $\Delta E = E_{\rm jet}$ and add $\delta(\Delta-E_{\rm jet}) \int^\infty_{E_{\rm jet}} d\epsilon P(\epsilon)$. This procedure is known as non-reweighting \cite{EskolaUrs}.
We point out that the alternative concept of reweighting
to our understanding systematically overestimates $P(\Delta E)$ for $\Delta E \ll E_{\rm jet}$ and should be disregarded. In fact for a dense medium increasing $\hat{q}$ and employing reweighting leads to an increased escape probability whereas increasing $\hat{q}$ and non-reweighting leads to the expected decrease in escape probability, see also \cite{Blackness}. }

From the energy loss distribution given a single path, we can define the averaged energy loss probability distribution for a given angle $\phi$ as
\begin{equation}
\label{E-P_phi}
\langle P(\Delta E)\rangle_\phi \negthickspace = 
\int_{-\infty}^{\infty} \negthickspace \negthickspace \negthickspace \negthickspace dx_0 
\int_{-\infty}^{\infty} \negthickspace \negthickspace \negthickspace \negthickspace dy_0 P(x_0,y_0)  
P(\Delta E)_{path}
\end{equation}
(this is conceptually similar to the angular averaged distribution $P(\Delta E)\rangle_{T_{AA}}$ introduced in \cite{Gamma-Tomography} for central collisions).

We calculate the momentum spectrum of hard partons in leading order perturbative QCD (LO pQCD) (explicit expressions are given in \cite{Correlations} and references therein). The medium-modified perturbative production of hadrons at angle $\phi$ can then be computed from the expression
\begin{equation}
\frac{d\sigma_{med}^{AA\rightarrow h+X}}{d\phi} \negthickspace \negthickspace = \sum_f \frac{d\sigma_{vac}^{AA \rightarrow f +X}}{d\phi} \otimes \langle P(\Delta E)\rangle_\phi \otimes
D_{f \rightarrow h}^{vac}(z, \mu_F^2)
\end{equation} 
with $D_{f \rightarrow h}^{vac}(z, \mu_F^2)$ the fragmentation function with momentum fraction $z$ at scale $\mu_F^2$ \cite{KKP}, and from this we compute the nuclear modification function $R_{AA}$ vs. reaction plane as
\begin{equation}
R_{AA}(p_T,y,\phi) = \frac{dN^h_{AA}/dP_Tdy d\phi}{T_{AA}({\bf b}) d\sigma^{pp}/dP_Tdy d\phi}.
\end{equation}

\begin{figure}
\includegraphics[width=0.9\linewidth]{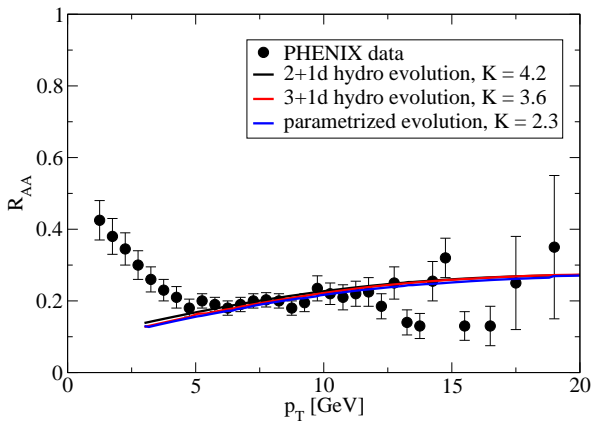}
\caption{$R_{AA}$ for central collisions as calculated in three different models for the medium evolution \cite{Nonaka:2006yn,Hydro,Parametrized} with the overall quenching power scale $K$ adjusted to data. 
}
\label{fig1}
\end{figure}

In \cite{Correlations,Gamma-Tomography} it has been shown that $R_{AA}$ 
for central collisions only constrains a scale, but not the detailed 
functional form of $\langle P(\Delta E) \rangle_{T_{AA}}$. In the approach 
outlined above, this is manifest in the parameter $K$ which we adjust 
to the data in central collisions. We illustrate in Fig.~\ref{fig1} that 
three different dynamical models, a 2D hydrodynamical evolution \cite{Hydro}, 
the 3D hydrodynamical evolution outlined above \cite{Nonaka:2006yn} and a 
parametrized fireball evolution \cite{Parametrized} give almost equal 
descriptions of $R_{AA}$ once the scale parameter is adjusted, albeit they 
require different values of $K$ (the chief reason for this being the 
different longitudinal dynamics). 
\jr{All three dynamical models provide a successful description of the bulk properties of the medium at RHIC
in central collisions \cite{Hydro,Nonaka:2006yn,Parametrized}. The parameterized evolution is adjusted in such a way that it also describes the Hanbury-Brown Twiss (HBT) correlation measurements correctly.
Given this successful description of measured observables within the three evolution models} the $\pm 50\%$ spread in the values of K
for the different models of the medium \jr{can therefore be} interpreted as a measure for the systematic 
error inherent in the tomographic analysis of jet energy-loss via the
nuclear modification function $R_{AA}$.

However, one may gain predictive power in going 
to collisions at finite impact parameter ${\bf b}$. 
The $\phi$ dependence 
of $R_{AA}$ for non-central collisions constitutes a systematic variation 
of path-length within a system with fixed overall scale.

\jr{Hydrodyamical models as \cite{Nonaka:2006yn} are able to provide the best framework for studying collisions at finite impact parameter whereas the application of a parameterized evolution model as \cite{Parametrized} to non-central collisions encouters the 
difficulty of how to implement eliptic flow appropriately. In the following we therefore exclusively resort to the 3D hydrodynamical evolution \cite{Nonaka:2006yn} which provides a very successful framework for the description bulk properties also for non-central collisions.}

The average 
path-length is expected to be smaller for a parton emitted in plane as 
compared to one emitted out of plane, and hence $R_{AA}$ is expected 
to be larger at $\phi=0$ than at $\phi=\pi/2$ with the difference in 
$R_{AA}$ between these angles increasing with the initial asymmetry 
(and hence ${\bf b}$). Using a simple model
for the time-evolution of the medium and collective flow effects, it
has been shown in \cite{Majumder:2006we} that the $\phi$ dependence
of $R_{AA}$ is quite sensitive to the initial gluon density distribution and
temporal evolution of the medium.

\begin{figure}
\includegraphics[width=0.9\linewidth]{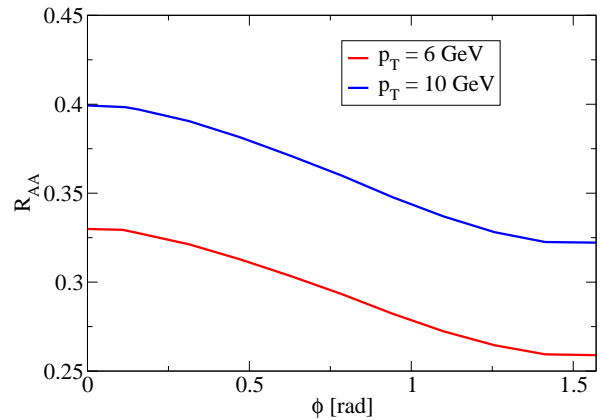}
\caption{Angular dependence of $R_{AA}$ for ${\bf b} = 7.5$ fm for two different values of $p_T$.
}
\label{fig2}
\end{figure}

\begin{figure}
\includegraphics[width=0.9\linewidth]{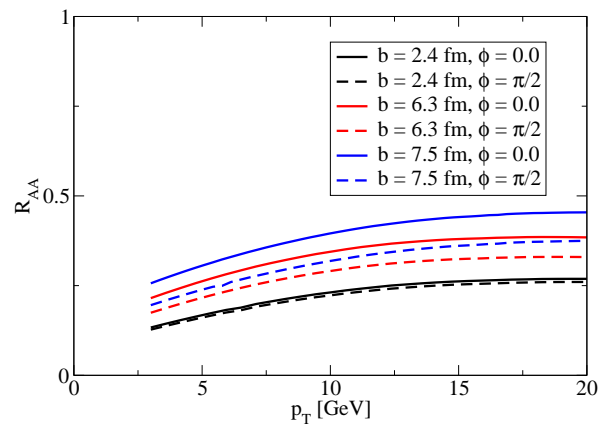}
\caption{$p_T$ dependence of $R_{AA}$ in plane (solid) and out of plane (dashed) emission at different values of impact parameter ${\bf b}$.
}
\label{fig3}
\end{figure}

Utilizing the previously discussed 3-D RFD model \cite{Nonaka:2006yn}, we study the angular dependence of $R_{AA}$ for two fixed values of 
$p_T$ at ${\bf b} = 7.5$ fm in Fig.~\ref{fig2} and show the $p_T$ 
dependence of $R_{AA}$ for emission in plane and out of plane at 
three different impact parameters ${\bf b}$ in Fig.~\ref{fig3}.

\begin{figure}
\includegraphics[width=0.9\linewidth]{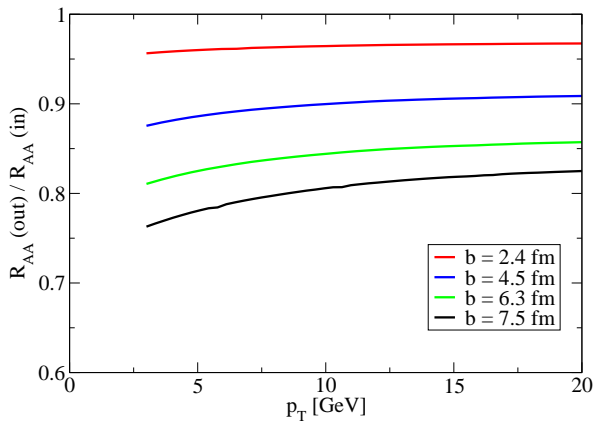}
\caption{Ratio of $R_{AA}$ for out of plane vs. in plane emission as a function of $p_T$ for four different values of the impact parameter ${\bf b}$.
}
\label{fig4}
\end{figure}

As expected, $R_{AA}$ grows for more peripheral collisions as there is less soft matter produced to induce energy loss. Moreover, there is a smooth angular variation of $R_{AA}$ observed, reflecting the underlying medium asymmetry. The difference between in-plane and out of plane emission grows with impact parameter, at ${\bf b} = 2.4$ fm there is hardly angular variation whereas at 7.5 fm differences are of order 20\% (see Fig.~\ref{fig4}).

\begin{figure}
\includegraphics[width=0.9\linewidth]{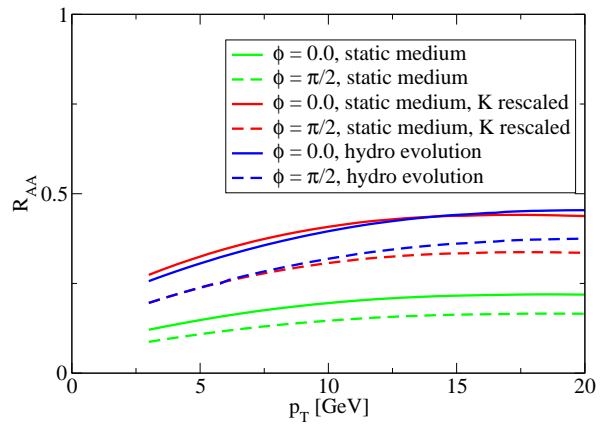}
\caption{$R_{AA}$ for in plane and out of plane emission as a function of $p_T$ at {\bf b} = 7.5 fm, assuming a static medium given by the hydrodynamical initial state, a static medium with readjusted quenching power given by the hydrodynamical initial state and the full hydrodynamical evolution.
}
\label{fig5}
\end{figure}

It is clear from the observation of elliptic flow that the pressure of the hydrodynamical fluid tends to remove the initial spatial asymmetry from the almond-shaped overlap region in non-central collisions. Thus, on general grounds, we may expect that in a dynamical model for the medium, the difference between in plane and out of plane emission is less pronounced than in an estimate using a static medium distributed according to Eq.~\ref{E-Profile}. On the other hand, the main fraction of observed hadrons arises from vertices close to the medium surface and the expected energy loss per unit time $\frac{d\Delta E}{d\tau}$ reached a peak value early in the evolution \cite{Correlations}, thus it is reasonable to expect that a large number of partons escapes the medium before the spatial asymmetry is completely removed. We investigate this competition of timescales in Fig.~\ref{fig5} where we make the comparison with a scenario in which we keep the medium static at its initial value.

Not altering the value $K=3.6$ for the quenching scale, keeping the energy density at its value in the initial state vastly overpredicts the quenching power of the medium, leading to $R_{AA}$ of the order of 0.25 for ${\bf b} = 7.5$ fm. Thus, we have to readjust $K$ to account for the (unphysical) fact that we keep the medium static. $K =  0.65$ leads to a good description for central collisions, and employing this value also at ${\bf b} = 7.5$ fm allows a fair assessment of the effect of the hydrodynamical evolution. 
\jr{The azimuthal dependence of $R_{AA}$ in the BDMPS-formalism without considering expansion effects using a fixed energy density profile has also been studied in \cite{Dainese}. }

One finds that indeed the difference between in plane and out-of plane emission is reduced by some 7\% when the hydrodynamical evolution is taken into account properly. In addition, a small change of the shape of the distribution with $p_T$ is induced. Thus, the model shows sensitivity to the size of the spatial asymmetry in the distribution of matter and the timescale at which it is removed. This complements the information found in low $p_T$ $v_2$ which reflects the asymmetry in momentum space.

\begin{figure}
\includegraphics[width=0.9\linewidth]{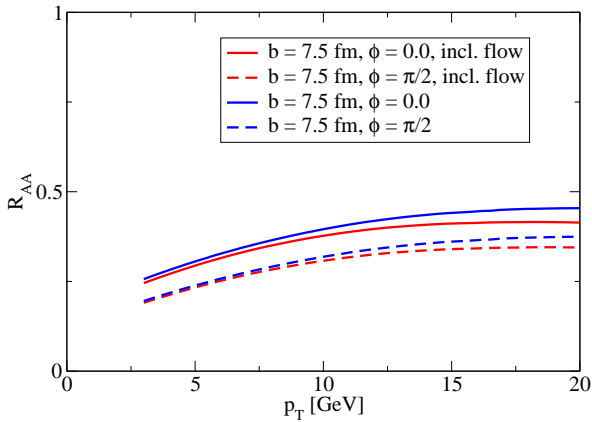}
\caption{$R_{AA}$ for in plane and out of plane emission as a function of $p_T$ for {\bf b} = 7.5 fm, comparing the result with and without explicit dependence on flow.
}
\label{fig6}
\end{figure}

However, jet energy loss can also couple to collective flow \cite{Urs2,JetFlow}, thus potentially blurring the relation to the spatial asymmetry. In order to assess this effect, we implement the effect of flow based on an expression based on the definition of $\hat{q}$ in AdS/CFT. We scale

\begin{equation}
\hat{q}' = \hat{q} (\cosh \rho - \sinh \rho \cos\alpha)
\end{equation}

where $\alpha$ is the angle between flow and hard parton trajectory and $\rho$ is the flow rapidity \cite{UrsPrivate}. This expression has a straightforward interpretation in terms of the density of scattering partners seen by the hard parton per unit time/length from the c.m. frame. The resulting effect on $R_{AA}$ at ${\bf b} = 7.5$ fm is shown in Fig.~\ref{fig6}.
\jr{Note that the scenario with transverse flow requires an increase
of $K$ by about $30 \%$ to describe $R_{\rm AA}$ in central collisions, in agreement with \cite{Schiff} transverse flow indeed decreases the quenching power.}

Transverse flow is in general weak during the timescale of jet quenching (i.e. during the early evolution), thus there is no pronounced difference between the curves. While we observe some change in shape, there is no significant change in the asymmetry between in plane and out of plane emission.

Let us now make a rough comparison with the preliminary PHENIX data \cite{PHENIX_data} (since systematic errors due to the reaction plane resolution are large, we refrain from doing a detailed comparison at this point). While the measurement appears to be consistent within errors with the calculation for $p_T > 6$ GeV, the data show a trend towards a greater angular spread between in plane and out of plane emission than found in the calculation. If this trend is confirmed, it would indicate that parametrically the path-length dependence of energy loss is larger than $L^2$ and would clearly rule out a linear dependence on $L$ (as expected e.g. for elastic collisions as the dominant mechanism for energy loss). While the $L^2$ dependence can be understood from coherence length phenomena (see e.g. \cite{Salgado:2003gb}), a dependence on a higher power of $L$ has so far no theoretical explanation. The apparent disagreement between calculation and data for $p_T$ below 6 GeV is yet another hint that fragmentation is not the dominant mechanism for the production of hadrons in this momentum regime and other mechanisms, such as parton recombination \cite{Reco} are of increasing importance for lower $p_T$.

In summary, we have discussed jet energy loss within a 3D hydrodynamical description of the medium. We have made no approximation in computing the average of the energy loss probability $\langle P(\Delta E) \rangle_\phi$ over all possible paths through the medium (in particular, we have not tried to solve the problem by identifying a typical path). 

Since the nuclear suppression factor $R_{AA}$ is not sensitive to the detailed form of $\langle P(\Delta E) \rangle_\phi$ as long as the quenching scale is adjusted correctly, using a hydrodynamical medium does not improve the quality of the description of $R_{AA}$ {\it per se}. However, once we fix the quenching scale to $R_{AA}$ in central collisions, we gain predictive power when going to more peripheral collisions. In particular, studying the angular dependence of $R_{AA}$ with respect to the reaction plane gives systematic control over the average in-medium path-length. 

As expected, the angular dependence of $R_{AA}$ reflects the spatially asymmetric distribution of soft matter. We also observe that the fact that the hydrodynamical evolution removes the asymmetry results in a reduction of the angular spread as compared to a situation in which the asymmetry is kept unchanged. We have also gauged the potential impact of the flow field on the results. The calculation describes preliminary data within errors, however there is a trend that the data show larger angular splitting than the calculation, indicating that the energy loss may scale parametrically with a larger power of the path-length than $L^2$. It remains to be seen if this trend can be confirmed.

{\it Acknowledgments:} This work was supported in part by an Outstanding Junior Investigator
Award  from the U.~S.~Department of Energy (grant DE-FG02-03ER41239) and
by the Academy of Finland, Project 206024. JR acknowledges support by the Natural Sciences and Engineering Research Council of Canada.

\end{document}